\documentclass{aa}  

\usepackage{graphicx}
\usepackage{color}
\usepackage{txfonts}
\begin{document}

   \title{Small impacts on the giant planet Jupiter}


   \author{R. Hueso                    \inst{1} \fnmsep\thanks{\email{ricardo.hueso@ehu.es}} \and
               M. Delcroix                 \inst{2} \and
               A. S\'anchez-Lavega      \inst{1} \and
                S. Pedranghelu \and
               G. Kernbauer    \and
               J. McKeon          \inst{3} \and
               A. Fleckstein      \and
               A. Wesley           \inst{4} \and
               J.M. G\'omez-Forrellad  \inst{5} \and
               J.F. Rojas                     \inst{1} \and
               J. Juaristi            \inst{1} 
             }

   \institute{F\' isica Aplicada I, Escuela de Ingenier\'ia de Bilbao, UPV/EHU, Alameda Urquijo s/n, 48013, Bilbao, Spain\\     
   \and  Societ\'e Astronomique de France, France\\ 
   \and  Meath Astronomy Group, Dublin, Ireland\\ 
    \and Astronomical Society of Australia, 1502 Rubyvale Rd, Rubyvale Queensland 4702 Australia\\  
   \and  Fundacio Observatori Esteve Duran, Spain }
 
   \date{\today}
 
  \abstract
   {Video observations of Jupiter obtained by amateur astronomers over the past
eight years have shown five flashes of light with durations of 1-2 s. The first three of these events occurred on 
3 June 2010, 20 August 2010, and 10 September 2012. Previous analyses of their 
light curves showed that they were caused by the impact of objects of 5-20 m in diameter,
depending on their density, with a released energy 
comparable to superbolides on Earth of the class of the Chelyabinsk airburst. The most recent two flashes on Jupiter were detected on 
17 March 2016 and 26 May 2017 and are analyzed here.} 
   {We characterize the energy involved together with the masses and sizes of the objects that produced 
these flashes. The rate of similar impacts on Jupiter provides improved constraints on the 
total flux of impacts on the planet, which can be compared to the amount of exogenic 
species detected in the upper atmosphere of Jupiter.}
   {We extracted light curves of the flashes and calculated the masses and sizes of the impacting objects after calibrating 
each video observation. An examination of the number of amateur observations of Jupiter as a function of time over the past years allows
us to interpret the statistics of these impact detections.}
   {The cumulative flux of small objects (5-20 m or larger) that
impact Jupiter is predicted to be low (10-65 impacts per year), 
and only a fraction of them are potentially observable from Earth (4-25 per year in a 
perfect survey). }
   {We predict that more impacts will be found in the next years, with Jupiter opposition displaced toward 
summer in the northern hemisphere where most amateur astronomers observe. Objects of this size contribute negligibly to the abundance of exogenous species and dust in the stratosphere of Jupiter
when compared with the continuous flux from interplanetary dust particles punctuated by giant impacts. Flashes of a high enough 
brightness (comparable at their peak to a +3.3 magnitude star) could produce an observable debris field on the planet. We 
estimate that a continuous search for these impacts might find these events once every 0.4 to 2.6 years.}

   \maketitle
%

\section{Introduction}

Because of its large gravitational attraction and effective cross section, 
the giant planet Jupiter is the most likely place to receive impacts in the solar system. 
Direct and dramatic evidence of impacts on Jupiter was acquired with the observations 
of the series of impacts from the comet Shoemaker-Levy 9 (SL9) in July 1994 
\citep{harrington2004book}. On July 19, 2009, an unknown body collided 
with Jupiter on its night side \citep{SanchezLavega10}. In both cases, the impacts produced 
large spots of material that were dark in the visible wavelength range and bright 
in methane absorption bands because of the high altitude of the debris fields. These spots 
remained visible for weeks to years \citep{Hammel95, Hammel10, SanchezLavega98, SanchezLavega11}. 

Impacts supply disequilibrium species to the upper atmosphere of Jupiter, which in the case of SL9 are still 
unambiguously observable today because of the higher concentration of water, 
CO, and other chemical species in the southern hemisphere of Jupiter 
\citep{Lellouch02, Cavalie13}. Spectroscopic observations of the planet 
allow inferring the amount of exogenic molecules in the upper atmosphere of Jupiter. 
However, the relative contribution to the abundance of exogenic water and carbon dioxide from impacts of very 
different size range from a continuous supply of interplanetary dust to rare impacts of large objects 
are not yet well characterized \citep{Lellouch02, Bezard02, Poppe16, Moses17}.

Impacts from objects of about 10 m in diameter have been detected in telescopic observations of Jupiter 
from the sudden release of luminous energy when  the impacting objects enter the 
atmosphere of Jupiter and explode as atmospheric bolides \citep{Hueso10b, Hueso13}. 
Five impacts have been detected in this way since 2010. Three of them have been  
examined  previously in the literature \citep{Hueso10b, Hueso13}, and two more have occurred since then. 
Each of these impacts has been detected simultaneously by more than one observer (12 observers 
recorded 11 video acquisitions for a total of five impact bolides). 

\cite{Hueso10b} presented the analysis of the first bolide impact on Jupiter and the general method
for calibrating light curves from amateur observations of Jovian flashes. They also converted them 
into energies, masses, and sizes of the impacting object. The first Jovian bolide was caused by an object with a diameter 
in the range of 10 m, releasing an energy comparable to an object of about 30 m impacting Earth's atmosphere. 
Follow-up observations with telescopes such as the Very Large Telescope (VLT) or the Hubble Space Telescope (HST) 
showed no evidence of atmospheric debris left by the impact, 
confirming the small size of the object. \cite{Hueso13} extended this study to a characterization of the three known 
bolides in 2013. The combined analysis of these three impacts allowed a first quantification of the flux of similar impacts 
on Jupiter. The expected number was 12-60 impacts per year for objects larger than 5-20 m in diameter. 
This impact rate is close to expectations based on an extrapolation of dynamical models of comets and asteroids 
in orbits prone to Jupiter encounters \citep{Levison00} (30-100 collisions per year 
for objects with diameters larger than 5-20 m). 
\cite{Hueso13} also presented model simulations of  airbursts caused by these small-sized objects following similar techniques to those used in simulations of larger impacts 
\citep{kory2006apj, palotai2011apj, pond2012apj}. 

We here update previous results presented in \cite{Hueso10b, Hueso13} with the analysis of the 
latest two impacts detected in March 2016 and May 2017. We examine how the new observations 
constrain the flux of impacts on Jupiter similarly to the estimate in \cite{Hueso13}. 
We also discuss the implications of the predicted flux of impacts on the amount 
of exogenic species (water and carbon monoxide) and dust in the upper atmosphere of Jupiter, and we discuss
the probability of finding more intense flashes from larger objects that could leave observable
traces in the atmosphere of Jupiter from follow-up observations.

The outline of this paper is the following: In section 2 we summarize previous observations 
of the first three impact fireballs \citep{Hueso10b, Hueso13}, and we give observational details of the 
latest two impacts found on Jupiter in March 17, 2016, and May 26, 2017.
In section 3 we present light curves of these two impacts and calibrate the images to obtain size 
and mass estimations of these objects. In section 4 we examine the sizes and masses 
of impacts required to leave an observable debris field in the planet atmosphere. In section 5 we discuss current efforts
of detecting new observation flashes on Jupiter. In section 6 we present an statistical analysis of 
the amateur observations to infer the statistical significance of these impact detections in a larger context. 
In section 7 we present an updated estimate of the impact  flux on Jupiter and discuss 
the implications for exogenous water and carbon monoxide on the upper atmosphere  of Jupiter
and the probability of finding observable debris fields in the atmosphere of Jupiter after more intense 
superbolides. We present our conclusions and a summary of our findings in section 8.

\section{Observations}

\subsection{Summary of previous impacts }

   \begin{figure}[htp]
   \centering
   \includegraphics[width=7.50cm]{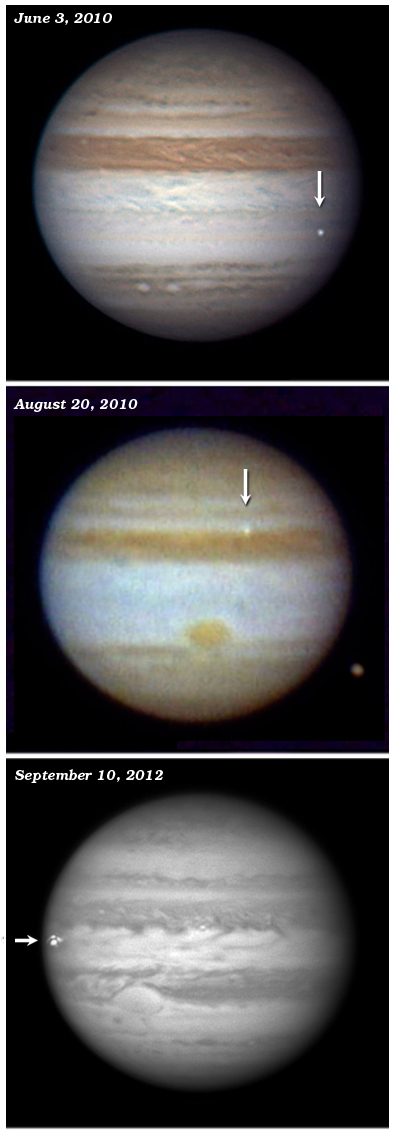}
   \caption{First three fireballs found on Jupiter from 
observations by A. Wesley (top panel), M. Tachikawa (middle panel), and 
G. Hall (bottom panel). Arrows show the position of the impacts. Images have been processed by stacking the frames 
where the flash is visible and adding the result over a Jupiter image built from the stack of the full 
video of the planet at the time of the impact. The color in the first panel comes from 
acquisitions obtained with a filter wheel in the minutes before
and after the flash. 
The color in the second panel comes from the detector, which makes use
of a Bayer mask to produce color images. The diffraction-like ring patterns 
in the last panel show the brightest flash.}
   \label{fig:Impacts1}
    \end{figure}

On June 3, 2010, Anthony Wesley (Australia) and Christopher Go (Philippines) 
recorded a short flash while taking video observations of Jupiter. The flash lasted about 
two seconds and was observed by A. Wesley, who issued an e-email alert that was confirmed later 
by C. Go. A quick and large follow-up campaign was organized, and it obtained observations of the planet 
within a few days with telescopes such as the VLT and HST. None of these observations found any debris 
field in the region hit by the impact. A later analysis of the flash light curve resulted in the conclusion 
that it was caused by an object  of 8-13 meters in diameter impacting the atmosphere of Jupiter and producing a 
giant fireball \citep{Hueso10b}. 

Two months later, on August 20, 2010, another flash on Jupiter was detected by amateur astronomer 
Masayuki Tachikawa of Japan, and it was confirmed by  Kazuo Aoki and 
Masayuki Ishimaru. This flash was found in videos of lower quality in the wake of published news
in amateur astronomy journals about the first flash. Two years later, on September 10, 2012, a new 
fireball on Jupiter was discovered by a visual observer, who issued an alert in astronomical 
forums (Dan Petersen from Racine in Wisconsin). The flash was later confirmed by a video observation 
obtained by George Hall from Dallas, Texas. This flash was significantly brighter than the previous two 
impacts. Subsequent analysis of the video observation agreed remarkably well with the brightness 
estimate from Dan Petersen, who had observed the flash 
visually in the telescope eyepiece.

Figure ~\ref{fig:Impacts1} shows processed versions of the observations of the three flashes. 
In these three cases (as well as in the next two flashes discussed
below), an observer raised the alert to 
the amateur community after observing the impact flash. Confirmations 
from other observers who had been taking data at the same time quickly followed, but we remark
that most observers did not see the flash originally when they
were at the telescope, or when they first analyzed their video observations with automatic stacking 
software tools. Amateur astronomers combine thousands of frames from a single video into a stacked 
image with high signal-to-noise ratio using automatic software tools \citep{Mousis14}, where the light of 
any possible short flash dilutes within the rest of the frames, rendering it invisible in the final image.
In all cases, the regions that were hit did not show any trace of the bolide material 
in later observations either by large telescopes such as in June 2010 and September 2012 or by 
fast amateur follow-ups in August 2010. Therefore these events can only be discovered if it is spotted  
in the few seconds during which each impact produces a bright fireball.

   \begin{figure}[htp]
   \centering
   \includegraphics[width=7.50cm]{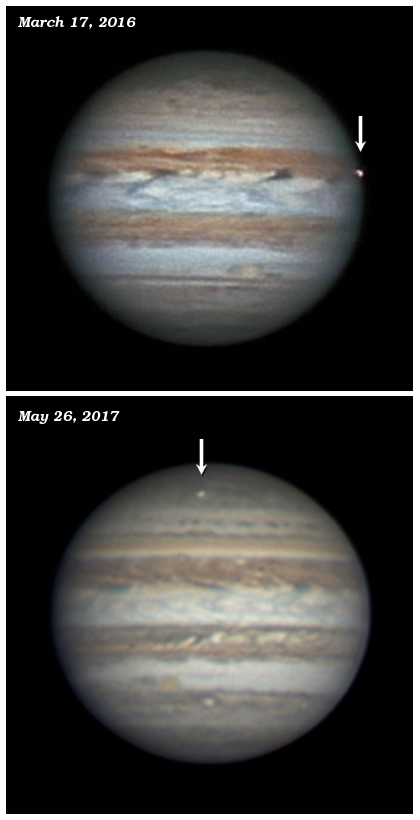}
   \caption{Latest two fireballs found on Jupiter. Arrows show the position of the impacts. 
Upper panel: Impact on March 17, 2016, observed by 
G. Kernbauer and J. McKeon. The image has been processed for aesthetics by 
S. Voltmer using data from the two video observations. The color 
comes from the G.K. video and the luminance from J. McKeon. Bottom panel: Impact on May 26, 2017,
observed by S. Pedranghelu, T. Riessler, and A. Fleckstein. The image is a 
combination of video observations in color from S.P. and T.R. All image combinations were done with stacking software 
and wavelet processing to increase the sharpness of atmospheric features, and the impact was added from a separate 
processing of the frames where the flash is visible.}
   \label{fig:Impacts2}
    \end{figure}

\subsection{Impact on March 17, 2016}

A new impact was detected on March 17, 2016, by Gerrit Kernbauer (Austria). 
The impact was announced ten days after the observation because the relatively poor seeing of that night caused
G.K. to delay an analysis of the video observations. 
The announcement was noted by John McKeon (Ireland), who had been observing the planet in the same night  for 
3.5 hours, building a time-lapse video of Jupiter and its moons. The second video confirmed the finding with 
better image quality.  It is difficult to visually find a short flash of light of 2 seconds 
in a sequence of video observation that lasts several hours.

\subsection{Impact on May 26, 2017 }
The most recent impact on Jupiter was found by Sauveur Pedranghelu from 
Corsica (France) on May 26, 2017. The impact was announced the next day 
and was quickly confirmed by two German observers, Thomas Riessler and Andr\'e 
Fleckstein, both after reading news of the impact posted on German astronomical 
forums. The videos by S.P. and T.R. were of excellent quality, and the video observation 
by A.F. was not as good because of the atmospheric seeing at his location. It was almost impossible to observe the flash in
this video without previous 
knowledge of the moment where the impact had occurred, meaning that good 
atmospheric seeing is a critical factor in the discoveries of these events. 
Figure  ~\ref{fig:Impacts2} shows these two impacts from a variety of video observations. 
Table  ~\ref{tab:obs} summarizes the dates of all the impacts, observers, and equipment.

\begin{table*}[]
\caption{Jovian bolides detections}
\centering 
\begin{tabular}{c l c c c c c}
\hline
\hline 
Date (yyyy-mm-dd)        & Observers             & Telescope   & Detector  & Filters  & Flash        & Sampling \\
Time (hh-mm-ss)           & (and locations)      &  diameter   &               &            & duration   & rate \\  
(UT)                             &                            &  (cm)        &                &            & (s)          & (fps)\\  
\hline

2010-06-03                &  A.Wesley (Australia)       & 37          & Point Grey Flea3    & 650 nm & 1.9 &60       \\
(20:31:20)                 &  C. Go (Phillipines)          & 28          & Point Grey Flea3    & 435 nm & 0.95 &55       \\               
\\
2010-08-20                &  M. Tachikawa (Japan)    & 15          & Philips ToUCam II  & Bayer RGB    & 1.4  & 30         \\
(18:21:56)                 &  K. Aoki          (Japan)    & 23.5        & Philips ToUCam II  & Bayer RGB    & 1.9   & 15         \\
                                &  M. Ishimaru  (Japan)    & 12.5        & Philips ToUCam II  & Bayer RGB     & 1.1  & 30         \\
\\
2012-09-10                &  D. Petersen (USA)         &  30.5       & Visual observation   & ---         & ---- & --- \\
(11:35:30)                 &  G. Hall       (USA)          &  30.5       & Point Grey Flea3     & 640 nm  & 1.7 & 15   \\
\\
2016-03-17                &  G. Kernbauer (Austria)    & 20          & QHY5LII               & Bayer RGB      & 1.15 & 47     \\
(00:18:39)                 &  J. McKeon (Ireland)        & 28          & ASI120MM            & IR742            & 1.30 & 26      \\  
\\
2017-05-26                &  S. Pedranghelu (France)  & 20.3      & ASI224MC   & Bayer RGB        & 1. 38 & 61.79  \\
(19:24:50)                 &  T. Riessler (Germany)      & 20.3      & ASI120MC   &  Bayer RGB       & 0.88 & 30.78 \\
                                &  A. Fleckstein (Germany)  & 28         & ASI120MM  &  IR742              & 0.92    & 30     \\
\hline
\end{tabular}
\label{tab:obs}
\tablefoot{* fps stands for frames per second}
\end{table*}

\subsection{Follow-up observations}

For the 2016 and 2017 impacts, observations by a variety of amateur astronomers were obtained from between 10 minutes to a few 
rotations after the impact. None of these observations showed any debris field on the surface of the planet 
at the locations of each impact. We surveyed the Planetary Virtual Observatory and Laboratory (PVOL) \citep{Hueso10a, Hueso18} 
and the Association of Lunar and Planetary Observers (ALPO) Japan databases 
of amateur observations as well as several amateur astronomical forums in search of images covering these
areas. In the case of the March 2016 impact, the impact area was just disappearing behind the east limb, and the high-resolution
observation closest 
in time was obtained by Randy Christensen (USA) one Jovian rotation later (10 hr). 

\begin{figure}[htp]
   \centering
   \includegraphics[width=8.70cm]{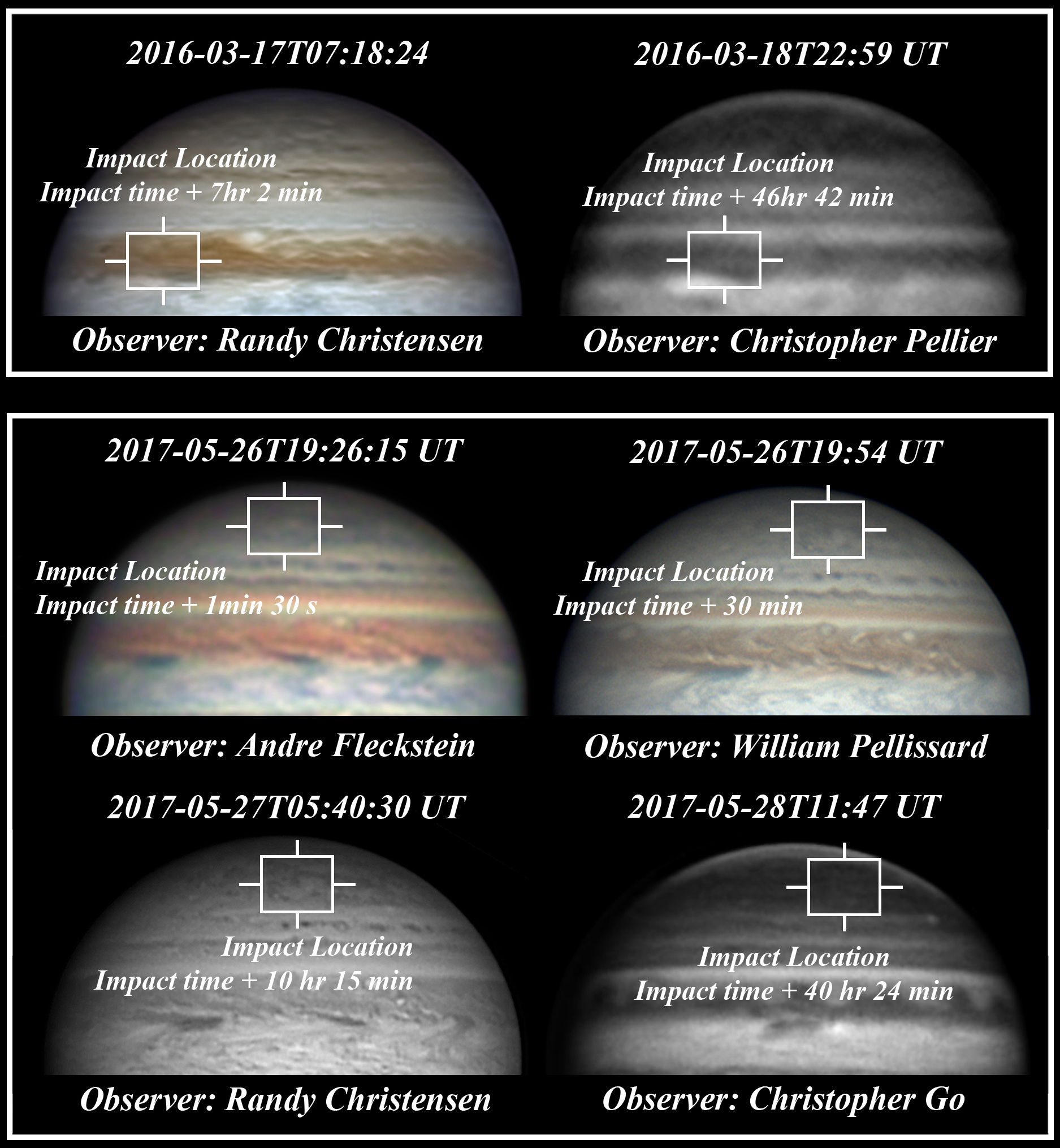}
   \caption{Selected follow-up observations of the events in March 2016 (top panel) and May 2017 (bottom panel).
In the follow-up observation by A.F., differential images with images acquired just before the flash
do not show any significant difference at the impact location.}
   \label{fig:follow-up}
    \end{figure}

For the May 2017 impact, the geometry was better suited. Images were acquired inmediately after the impact by 
A.F., but showed no brightening or a darkening of the impact area. Images with some better seeing 
were acquired 10 minutes  after the impact by Giancarlo Rizatto (Italy) and Philipp Salzgeberg 
(Austria) without any observable impact feature with a good image quality.  A better observation was obtained 
by William Pelissard (France) 30 minutes after the impact  and did not show an observable debris field. 
A high-resolution observation obtained  10 hours later by Randy Christensen did not show any observable perturbation 
at the impact location. Finally, a methane band observation by Christopher Go 40 hours after the impact did not show 
any bright feature in the planet. Figure \ref{fig:follow-up} shows a selection of these images. 


\section{Analysis of the impacts in March 2016 and May 2017}
   \begin{figure}[htp]
   \centering
   \includegraphics[width=8.50cm]{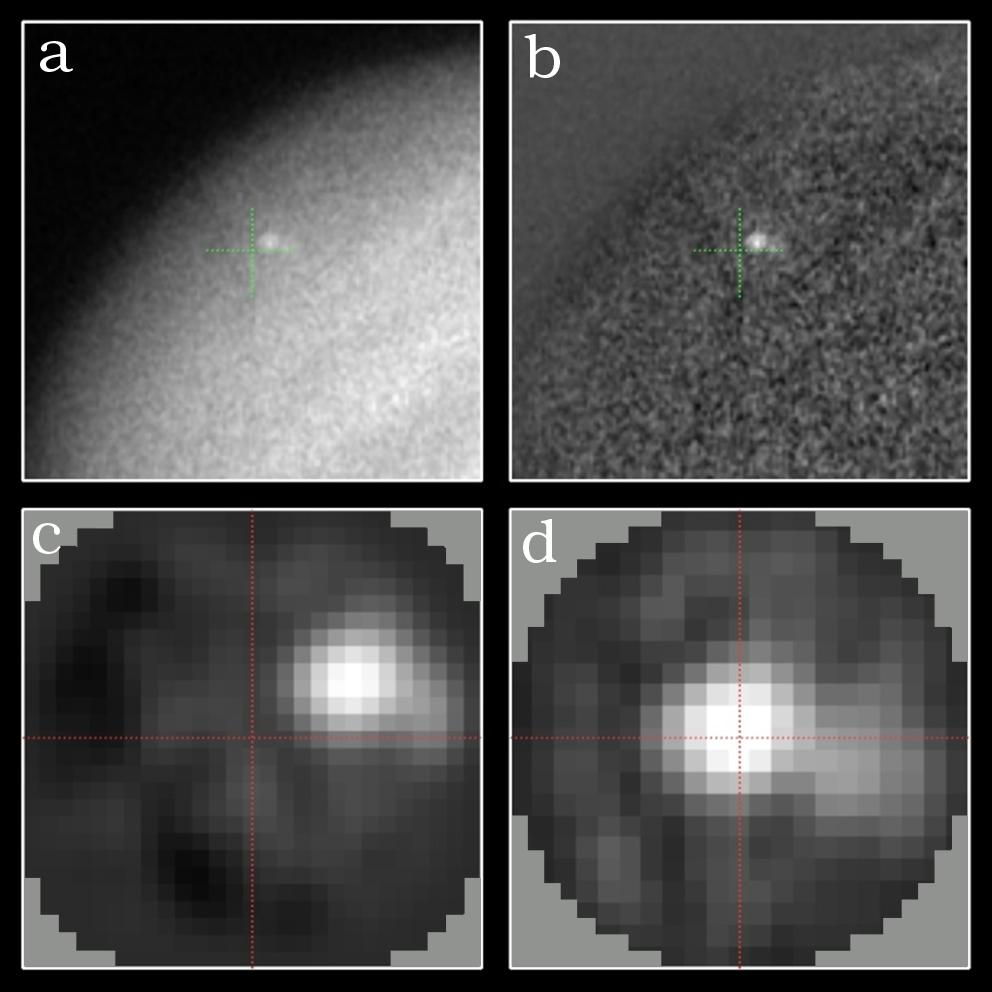}
   \caption{Light-curve analysis pipeline. (a) Section of the original frame; (b) differential 
image for that frame, i.e., the current frame minus the reference image; (c) circular mask around the mean
location of the impact; and (d) circular mask within a smaller radius containing all the light 
from the impact and recentered on the impact. Seeing distorts each image differently, and the location of the flash needs to be adjusted frame by frame by the software. 
The example is from  the S.P. video observation on 2017-05-26.}
   \label{fig:pipeline}
    \end{figure}

\subsection{Light curves}
For each video observation, we transformed the initial video files into a sequence 
of numbered frames that were analyzed with a software pipeline written in IDL. 
The pipeline coregisters all frames by calculating the relative motions of the frames caused
by the atmospheric seeing. A reference image is calculated from a stack of coregistered images and 
normalized by taking into account the number of frames used.
The coregistration is done by an image correlation algorithm and is loosely based on the 
PLAYLIST pipeline for lucky imaging of planets \citep{Mendikoa16}. The impact location is found 
by calculating an image built from the maximum brightness of each pixel and 
substracting the average brightness for each location. This generally produces an image 
where the impact location is well contrasted and can be found automatically. 
However, atmospheric seeing
not only moves the planet from one frame to another, it also distorts the planet shape, causing the 
flash light to apparently move around the main impact location. 

In order to calculate the light from the flash, the software calculates the difference between each frame
and the reference image after coregistering each individual frame with the reference image. 
Differential photometry images are used to calculate aperture photometry over the impact location.
Aperture photometry is done using a circular mask plus 
a ring to substract contributions from the background in the differential photometry images. 
The circular mask is recentered in each frame above the flash location by correcting distortions 
caused by the seeing. Additionally, the software also calculates the integrated flux of Jupiter so that the brigthness 
of the impact can be compared with the total brightness of the planet. Figure \ref{fig:pipeline}
shows examples of the pipeline processing. Images like these are generated by the pipeline and are 
used to check the correct positioning of the moving-aperture photometry mask. 

For video observations with cameras that use a Bayer RGB filter, we convert the color frames into 
black-and-white versions using the average of all three channels. This allows better visibility of the impact and a
more detailed and less noisy light-curve.

\begin{figure}
   \centering
   \includegraphics[width=9cm]{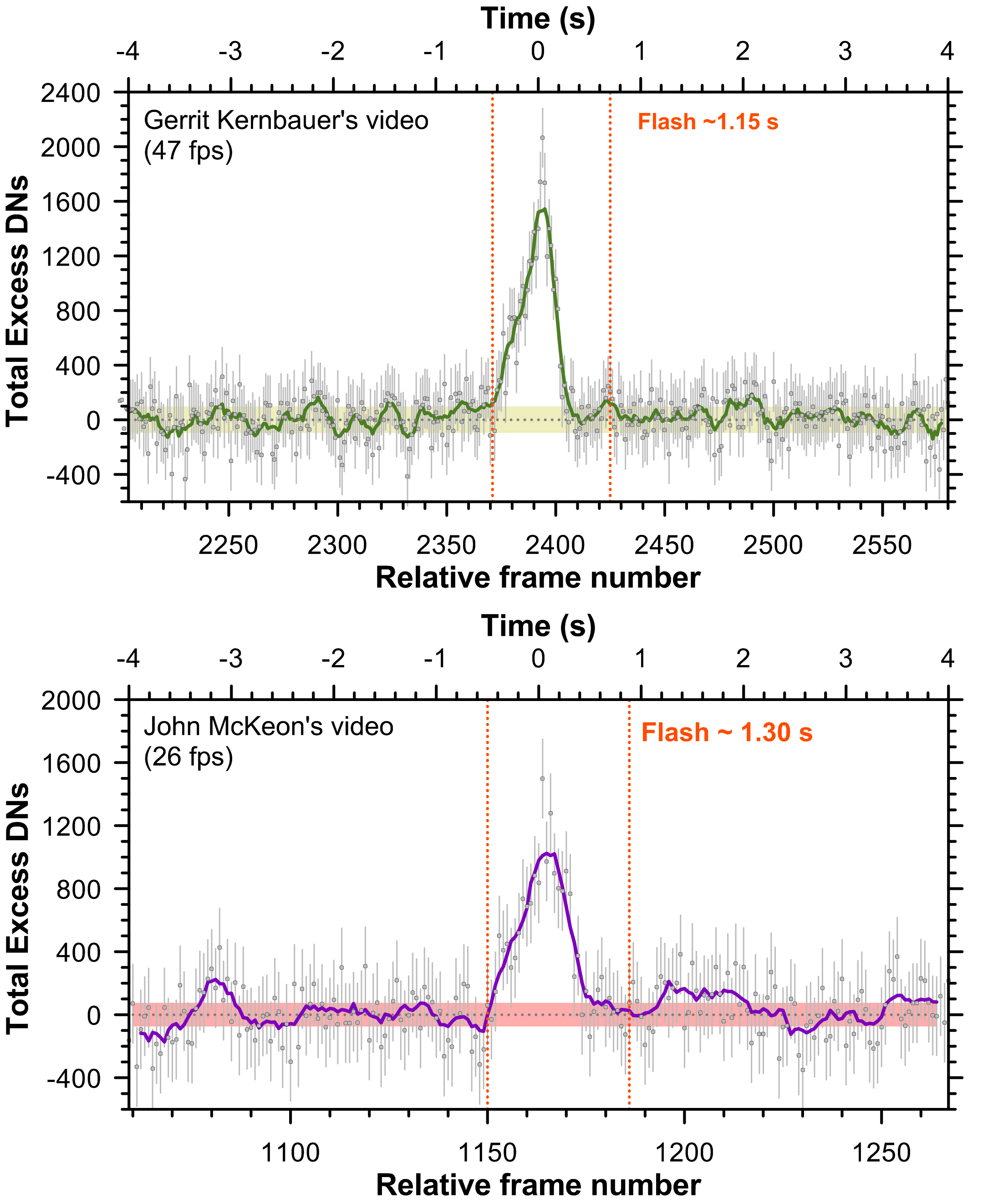}
   \caption{Raw light curves of the impact in March 2017. Top panel: Data from Gerrit Kernbauer. Bottom pannel: Data from John McKeon.}
   \label{fig:Impacts16}
    \end{figure}

\begin{figure}
   \centering
   \includegraphics[width=9cm]{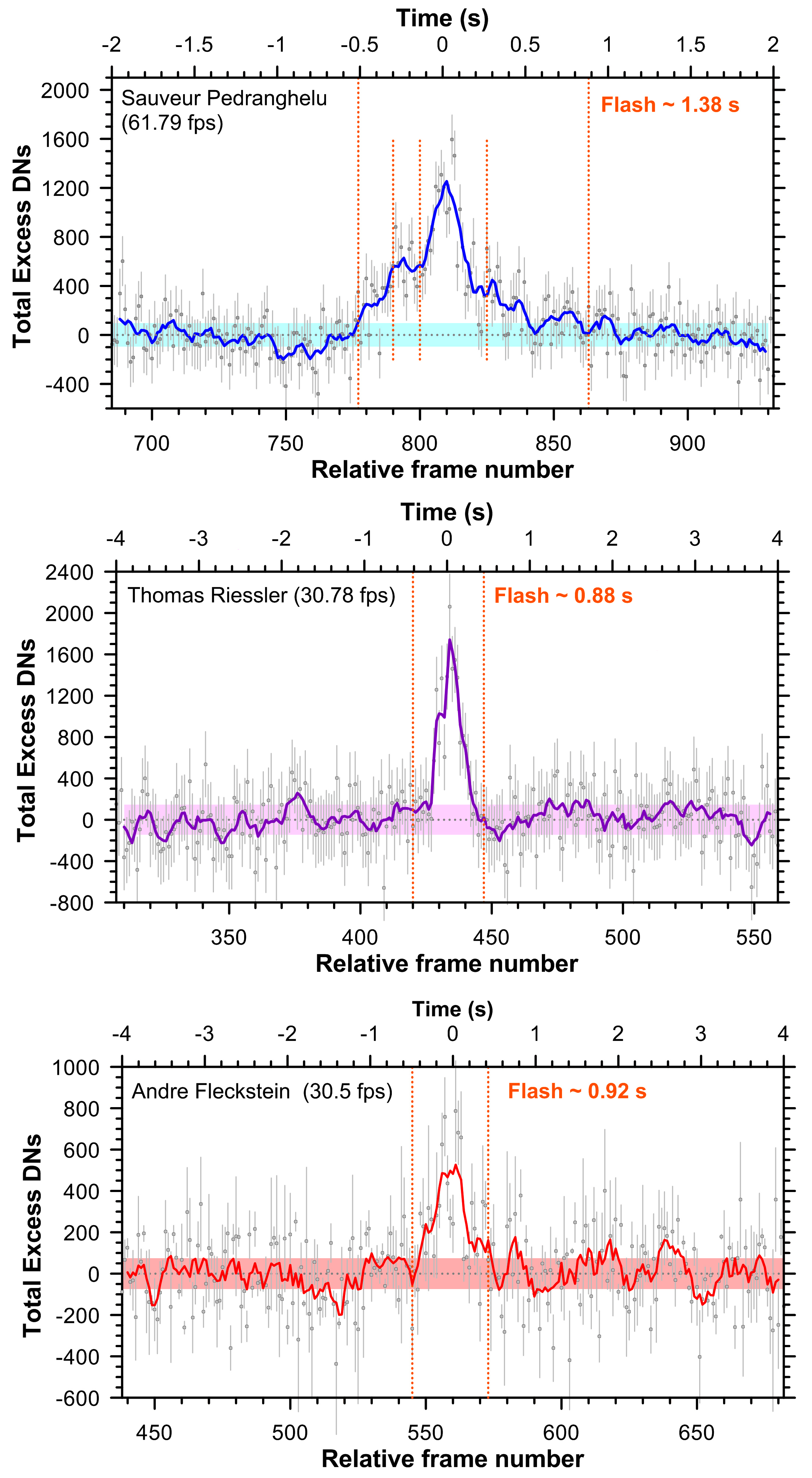}
   \caption{Raw light curves of the May 2017 impact. 
Top panel: Data from Sauveur Pedranghelu. The light curve has different phases that are identified 
with vertical orange lines: A first flash of 0.21 s, a second phase with constant flux for 0.16 s, 
a central flash of 0.40 s, and a extended decay for another 0.32-0.61 s for a total flash 
duration of 1.38 s. Middle pannel: Data from Thomas Riessler showing the double flash 
with a slightly shorter duration. Bottom panel: Data from Andr\'e Fleckstein. The video by 
T.R. does not show the same amount of structure visible in the first light curve, possibly 
because of the different frame rates and smaller pixel size. The video by A.F. was acquired 
under poorer seeing conditions.}
   \label{fig:Impacts17}
    \end{figure}

Figures \ref{fig:Impacts16} and \ref{fig:Impacts17} show 
raw light curves of the impacts in March 2016 and May 2017, respectively. In this case the light
curve from the analysis of S.P.'s video observation shows significant temporal structure with 
a double central flash and an extended tail of brighness decay lasting for about 0.6 s. The double central flash 
is also partially distinguishable in the second video of this event by T.R., but the smaller pixel size
of the optical setup prevents us from extracting more accurate information of this video. The two flashes are readily 
apparent when examining the video observation frame by frame and can be related to fragmentations 
of the impact object. The same types of features with similar timescales
are observed on light curves of superbolides on Earth (see, e.g., Figure 3 in \citealt{Borovicka17} and
the extended figure 2 in \citealt{Borovicka13}). This might therefore
be the
first case for fragmentation of Jupiter bolides and suggests that observations of future impacts
should involve a fast frame rate of at least 30 fps to observe these characteristics. 
The fragmentation history of a bolide depends on the entry mass, physical nature of the meteorite, 
speed, and angle of the impact. 
A sophisticated light curve analysis can be made for Earth superbolides resulting in the physical 
characterization of the impactor \citep{Borovicka17} (i.e., determining the physical 
class of the impacting object, which can be stony, metallic, icy compact, or icy porous). 
The fast cameras currently used by most amateurs 
might start to produce such data in observations of Jupiter impacts for objects more massive than the impact in May 2017.

   \begin{figure}
   \centering
   \includegraphics[width=9cm]{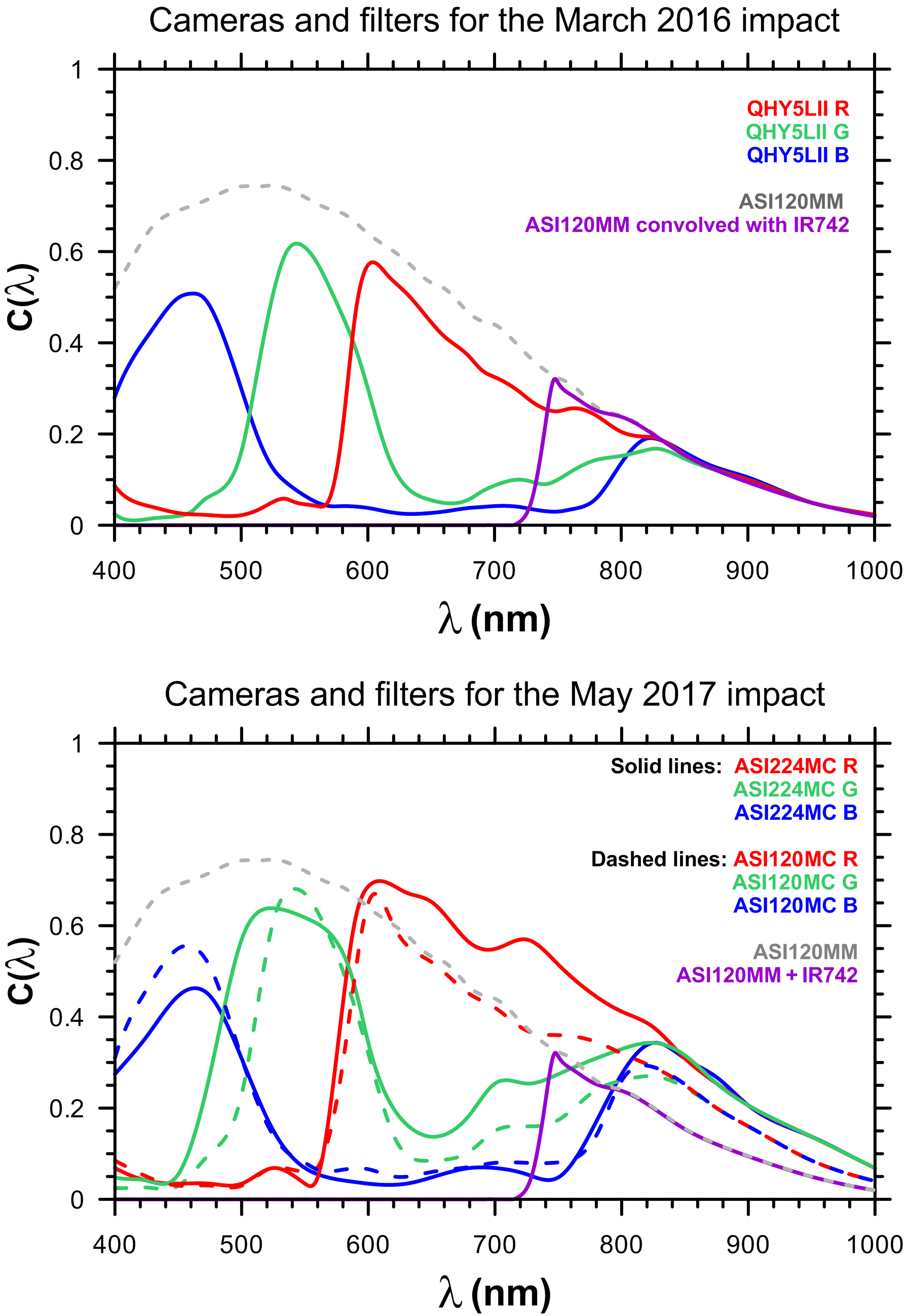}
   \caption{Spectral responses of the combination of cameras and filters for the impacts in March 2016 (top) and May 2017 (bottom). Note the better quantum efficiency of the 
ASI224MC camera used by S.P. for the May 2017 impact when compared with the ASI120MC used by T.R. and the QHY5LII camera used by G.K. for the impact in March 2016. All the color filters used correspond to Bayer mask filters on the CCD.} 
   \label{fig:QE}
    \end{figure}

\subsection{Image calibration}
The total intensity from each flash,  $I_{*}$,  is calculated from the integrated data numbers (DNs) of each light curve. This is computed by adding the excess DNs from the minimum to the maximum times marked in figures \ref{fig:Impacts16} and \ref{fig:Impacts17}.  This number is compared with the total DNs associated to the full disk of Jupiter over the reference image built for each video, $I_J$. Transforming $I_*$ into luminous energy, $L_*$ (measured in Joules), is a simple problem of scaling the flash light with the Jupiter brightness as detected in each observation,
\begin{equation}
L_*= K \cdot \left( \frac{I_{*}}{I_J} \right) , 
\end{equation}
where $K$ is a conversion factor different for each video observation and proportional to the total flux of light reflected from Jupiter and detected with the camera. 

The conversion factor was calculated in the following way: for each date, we calculated 
the effective solar constant at the distance of Jupiter, $S_J$, by scaling the solar constant 
at Earth, $S_E=1361$ $W/m^2$ to the distance of Jupiter to the Sun, $d_J$, using the
ephemeris computed with the JPL HORIZONS system at \url{https://ssd.jpl.nasa.gov/}. 
For each video observation we convolved the solar spectrum from \citet{Colina96} 
with the camera and filter response and the reflectivity spectrum of Jupiter from 
\citet{Karkoschka94} by computing the amount of energy reflected from Jupiter 
and detected by the camera. Because the impact photometry is computed 
with respect to the full disk brightness of Jupiter, the absolute values of the camera and 
filter response are not needed, only their relative values at different wavelengths.
Thus,

\begin{equation}
L_*= \left( S_J \cdot \pi R_{eq} R_{p} \right) \cdot  
\left(
\frac{\int_{0}^{\infty} F_\odot\ (\lambda)\cdot I/F(\lambda)\cdot C(\lambda) d\lambda}
{\int_{0}^{\infty} F_\odot\ (\lambda) d\lambda} 
\right)
\cdot \left( \frac{I_{*}}{I_J} \right)  
\cdot \Delta t,
\end{equation}
where $R_{eq}$ and $R_p$ are the equatorial and polar radius of the planet,
$F_{\odot}(\lambda) $ is the solar spectral radiance, $I/F(\lambda)$ is the reflectivity of Jupiter, $C(\lambda)$ is 
the spectral response of the camera and filter (figure \ref{fig:QE} 
shows the spectral responses of the cameras and filters used in the detection of the two impacts), and $\Delta t$ is the exposure time for a single frame.
The first term represents the flux of solar light intercepted by Jupiter. The second term is the
proportion of this energy that the system can detect and contains the spectrum of Jupiter and the spectral response of the camera. 
The third term contains the normalization factor from the analysis of the light curve and the integrated light of the Jupiter disk. 
The $\Delta t$ term is used to transform Watts into Joules. The result is the "detected" luminous energy of the flash given in Joules.

The flash behaves as a punctual source of light releasing energy in all directions. Part of this light (almost 50\%) 
illuminates the upper clouds of Jupiter and is reflected with an average approximate albedo of 0.5. This added contribution 
implies that a geometric correction needs to be introduced in the luminosity evaluation. The magnitude of the correction
depends on the viewing geometry of the impact, and its exact evaluation is a complex problem of radiative transfer.
For consistency with \cite{Hueso13}, this correction is here simply computed as 1.3,\begin{equation}
L_{*}^{cor}\approx L_*/1.3
.\end{equation}

This approximate correction is applied to the impact in May 2017, but not to the impact in March 2016, 
which occurred too close to the planet limb, to required adding this correction from reflections of light in the Jupiter clouds.

The "detected" and corrected luminous energy $L_{*}^{cor}$ corresponds to a part of the 
total emitted energy in form of light. Depending on the spectral energy distribution of this light,
the $L_{*}^{cor}$ can correspond to a higher or lower amount of total emitted light. If we assume that 
the light is emitted at a given temperature following Planck's blackbody law, we can compute
an efficiency factor for different temperatures  and detectors,

\begin{equation}
L_*^f(T)=L_{*}^{cor} 
\left( 
\frac{\int_{0}^{\infty} F_{BB}(T, \lambda) d\lambda} 
{\int_{0}^{\infty} F_{BB}(T, \lambda)\cdot C(\lambda) d\lambda}
\right),
\end{equation}
where  $L_*^f(T)$ is the total luminous energy of the impact as a 
function of temperature, $T$ is temperature, and $F_{BB}$ is 
Planck's law of radiation for a blackbody. We considered that 
blackbody brigthness temperatures of the flash are in the range 
of [3500-8500] K. These temperatures come from values of Earth's 
fireballs, SL9 impacts observed by the Galileo spacecraft, 
and analysis of the 2010 fireball on Jupiter, which was observed simultaneously 
at high quality with a red and blue filter \citep{Hueso10b}. 
This temperature range produces a factor of two uncertainty in 
the energy calculation, which is larger than the uncertainties in the light-curve calculation or the geometric factor correction.

In video observations obtained with cameras that use a Bayer mask to build RGB images, we 
considered that the image is the sum of the three red, green, and blue channel images, so that $C(\lambda)$ 
is the sum of the curves representing the spectral responses of each channel.

Finally, in order to transform the total luminous energy into kinetic energy of the impactor,
we need to know the efficiency of the impact to convert kinetic energy into luminous energy. 
For meteoroids and fireballs entering Earth's atmosphere, an empirical efficiency formula has been 
derived by \citet{Brown02}.

\begin{equation}
\mu=0.12 E_0^{0.115},
\end{equation}
where $E_0$ is the optical energy measured in kilotons of TNT (1 kton=$4.185\times10^{12}$ J). 

Values of $\mu$ from this formula for Jovian impacts range from 0.15 to 0.20.
We caution that this formula is calibrated from observations of Earth impacts with 
optical energies from 0.001 to 1 kiloton, while Jovian impacts release optical energies in the range of 5-60 kiloton.
Additionally, the impacts on Jupiter occur at a different velocity with an atmosphere of a different composition. 
These factors introduce an additional uncertainty in the size of the 
impact object that is currently unconstrained.


\begin{table*}[]
\caption{Jovian bolide analysis}
\centering 
\begin{tabular}{l c c c c c c c}
\hline\hline
Date           & Optical Energy & Kinetic Energy  & Energy & Mass                  & Diameter (m)                  & Diameter  (m)                 &  Diameter(m)\\
(yr-mm-dd) & (J)                  & (J)                  & (ktn)     & ($10^3$kg=ton)  & $\rho=2.0$ gcm$^{-3}$  &  $\rho=0.6$ gcm$^{-3}$  &  $\rho=0.25$ gcm$^{-3}$\\
\hline
2010-06-03*             &       $0.3-2.5 \times 10^{14}$  &   $1.9-14  \times 10^{14}$            &  46-350   &      105-780         &       4.7-9.1 &       7.0-14     &  9.3-18 \\ 
2010-08-20*             &       $0.6-2.0 \times 10^{14}$  &     $3.7-11  \times 10^{14}$        &       88-260   &      205-610         &         5.8-8.4 &       8.7-13     &  12-17  \\ 
2012-09-10*             &       $1.6-3.2 \times 10^{14}$   &         $9.0-17  \times 10^{14}$        &       215-405 &       500-950         &       7.8-9.7         &       12-14      &  15-19  \\
2016-03-17           &  $1.3-2.8 \times 10^{14}$  &     $7.3-14   \times 10^{14}$       &       175-350 &       403-805     &         7.3-9.2 &       10.9-13.7 &  14-19  \\  
2017-05-26          &   $1.9-3.6 \times 10^{13}$   &    $1.3-2.3  \times 10^{14}$     &  32-55    &      75-130       &  4.1-5.0 &       6.1-7.4     &  8.3-10 \\  
\hline
\end{tabular}
\begin{flushleft}
Note: (*) Data from \cite{Hueso13}. Densities of 0.25 $gcm^{-3}$ are considered as representative of the SL-9 impact \citep{Crawford97}.
\end{flushleft}
\label{tab:Analysis}
\end{table*}

\subsection{Masses and sizes of the impacting objects}
We assumed impacts at a velocity of 60 km s$^{-1}$ close to the 
escape velocity of Jupiter and densities from 2.0 to 0.25 g cm$^{-3}$.  The results 
are summarized in Table \ref{tab:Analysis} in comparison with 
determinations of energies and masses of previous impacts. 
The impact in May 2017 was approximately 4.5 times less
energetic than the impact in March 2016. 


When examining the ensemble of impacts on Jupiter given in Table \ref{tab:Analysis}, kinetic 
energies range 32-405 ktn close to Chelyabinsk-like events, which was 
considered to release about 450 ktn of energy \citep{Brown13} and an 
order of magnitude lower than the Tunguska impact (5,000-15,000 ktn) \citep{Boslough08}
or 1-3 million times lower than the combined SL9 impacts 
(estimated to release 300,000 kTn of energy) \citep{Boslough97}.

\section{Visibility of debris fields caused by intermediate-sized impacts}

The impacts we characterized have a remarkably small diversity of sizes. 
The largest of them was caused by an object of at most 19 m in diameter when 
considering a density of 0.25 $gcm^{-3}$ similar to the assumed density 
for SL-9 fragments \citep{Crawford97}. The smallest debris field left in the atmosphere 
of Jupiter by one of the SL-9 fragments was caused by fragment N. This fragment 
was estimated to have a size of about 45 m in diameter \citep{Crawford97}. 
For equal density, this is about 12 times more massive than the impact flash detected in 
September 2012, or 180 times more massive than the smallest impact flash 
in May 2017. The dark debris of fragment N was observed in HST images 
before it mixed with debris from other fragments  \citep{Hammel95}.

Fragment N would have caused a flash 12 times brighter than the 
impact in September 2012, producing significant saturation over a standard 
impact video record (most amateurs expose each frame,  so that the brightest 
part of Jupiter reaches about 70\% - 80\% of the saturation level in their detector). 
When we compare the estimated mass of fragment N and the impact
in 2012,
the video observation of that fragment would amount to 
a star of +3.3 magnitude. This is equivalent to three times the visible magnitude of 
Ganymede,  1/315 of the total flux of Jupiter, or $2\times10^{16}$ J. A flash of 
this energy would produce a debris in the atmosphere 
of Jupiter within the observable reach of HST and Earth's largest telescopes. 
Material from the N impact and other small SL9 impacts could be 
observed for at least two days, but there are no reports of them 
a week after the impact \citep{Spencer95}.

Dissipation times of impact material in the atmosphere for previous impacts 
were on the order of a few months for the 2009 impact \citep{SanchezLavega11} with
an e-folding time of~10 days in the debris particles concentration \citep{PerezHoyos12} 
and longer for the SL9 largest fragments \citep{SanchezLavega98}. In these impacts, the debris 
left in the atmosphere was spectrally dark in the continuum (with maximum contrast 
with the environment clouds at red wavelengths), but bright in methane absorption 
bands. This is so because the dark particles were deposited in the atmosphere at high altitude 
(pressures lower than 10 mbar)
\citep{Hammel95, Hammel10, dePater10, PerezHoyos12}. The impact debris was also 
bright in the thermal infrared because of the heating effect of an impact and the 
long radiative time constant of the stratosphere of Jupiter \citep{harrington2004book, dePater10}.

To the best of our knowledge, the long-term visibility of a small impact debris field has not yet been explored in the literature.
An empirical estimation between the debris lifetime and the size of the impactor would also depend
on the nature of the impactor (stony or icy), location of the impact in the atmosphere in a region
with higher or lower wind shear, and many other possible parameters (such as the impact trajectory angle with the planet).
However, in the event of an impact 10 times brighter than the flashing impact that occurred in September 2010, we 
predict that quick follow-up observations could detect an atmospheric debris field.

We examine the detectability of impact debris on Jupiter caused by larger impacts in section 7.3.

\section{Searches for new impacts}

\subsection{Dedicated detection campaigns}
Professional dedicated campaigns to detect impacts on Jupiter are difficult to carry out because
such campaigns must involve the capacity of acquiring and analyzing images over many 
different nights. Our team runs frequent observations of Jupiter with 1-2 m size telescopes using 
PlanetCam UPV/EHU, a lucky-imaging instrument \citep{SanchezLavega12, Mendikoa16}. 
In the past five years, we have observed Jupiter over 20 different campaigns over an average of 
two nights per campaign, acquiring about 2.0 hours of data for each night. These observations have 
been checked for impacts without observing an impact flash for an accumulated observing time of 80 hours. 
Details of the observations are given in \citet{Mendikoa16, Mendikoa17}.

The lack of impact detections in this survey imposes a weak constraint over
the maximum impact rate on the planet. We used a simple Monte
Carlo simulation to calculate the significance
of this negative detection. We tested different values of the number of observable impacts per year, and for each, we launched a large-number (5,000) of Monte Carlo simulations where we examined the number of impacts 
that would occur in 80 hours. To do this, we divided 80 hours into 4,800 opportunities of 1 minute to detect
an impact. We examined the statistics of the Monte Carlo simulation and searched for the number of observable 
impacts per year that result in 66\% of the simulations producing at least one impact in an accumulated
observing time of 80 hours. The statistical result
is that an impact rate of 120 detectable impacts per year would be needed to 
find one impact in this observing time. The Monte Carlo simulation shows 
that for a probability of 90\% to find an impact in  a survey of 80 hours, we would require 
an impact rate of 250 observable impacts per year. This analysis suggests an upper limit on the impact rate
smaller than 120 impacts per year.

A project run by Japanese amateurs called "Find Flash" and run by the 
Association of Lunar and Planetary Observers (ALPO) in Japan with more than 50 amateur observers
and about 10 nights per year observation time on 1 m size telescopes did not find impacts
for about four years of observations, placing a similar constraint (I. Tabe, private communication).

\subsection{Filters and technology}

Flashes can be best detected in filters where the planet is dark and the integration time of the camera
is not too long. Blue filters and relatively wide filters centered on the methane absorption band at 890 nm are 
best suited for a flash detection because the planet is dark and the flash should be bright. 
Integration times lower than 0.2 s are needed, suggesting that while blue filters can be used 
with small telescopes, an instrument with a minimum diameter of 30 cm might
be required to perform a flash detection campaign for the 890 nm methane band. 
Dual observations in blue and 890 nm would highly constrain the brigthness temperature 
of the flash, allowing for a determination of the impact energy with lower ambiguities.

The cameras used in the amateur community 
have experienced significant improvements over the past decade from read
noises of $\sim$ 8e- at 0 gain (e.g., in the popular Point Grey Flea3 camera
used for the first detections of impacts in Jupiter)
to $\sim$ 3e- at 0 gain (in the ASI cameras now used by most amateurs). Quantum efficiencies 
have improved at least by a factor of two in the near-infrared, where 
most observers concentrate their observational efforts, and have increased from 5\% to 30\% in the 900 nm range.
Better sensitivity results in impacts being detectable by smaller telescopes with 
faster exposures and better temporal resolution of the light curves.

\subsection{Impacts DeTeCtion software}
We have written a software tool to inspect amateur video observations of Jupiter
capable of detecting impacts in the planet. The latest version of this software, DeTeCt3.1, 
is an open-source application that amateur observers can use on their own computers \citep{Delcroix17}. 
The software constitutes one of the activities of the "Planetary Space Weather Services" (PSWS)
\citep{Andre18} and is based on differential photometry of coregistered images.
This software is essentially different from ''flash detection'' software developed for search of
impact flashes on the Moon \citep{Madiedo-MIDAS15} because these tools generally examine 
videos acquired over a large field of view (almost the full non-illuminated side of the Moon) without
the need to correct for seeing effects (coregistration) and because the flash is recorded over
a dark object and not the bright background of Jupiter.

The DeTeCt software has been used regularly by dozens of observers examining 
about 70,000 video files, which is equivalent to more than 75 days of observations distributed 
unevenly over several years. The software 
produces log files that are later analyzed to examine the statistics of nondetections when comparing 
with the fortuitous detections of impacts. The analysis shows that about 5\% of all observations were
acquired at the same time by the collaborating observers. The software and statistical analysis of its results can be accessed at
\url{http://www.astrosurf.com/planetessaf/doc/project_detect.shtml}.  
A similar analysis of video observations of Saturn is also available at that web site, with negative results so far.

From the statistics of the time covered by these observations, and because one of  
the observers of impacts in Jupiter (J. McKeon)  uses this software regularly, one can consider
a detectable impacts rate of 1/75 days$^{-1 }$), which is equivalent to 4.9 observable impacts per year.
This is a minimum number because most video files are acquired with regular sky conditions, and 
we cannot measure the eficiency of finding impacts on low-quality video files. 

\section{Statistical analysis of the impacts }

\begin{figure*}
\centering
\includegraphics[width=18.0cm]{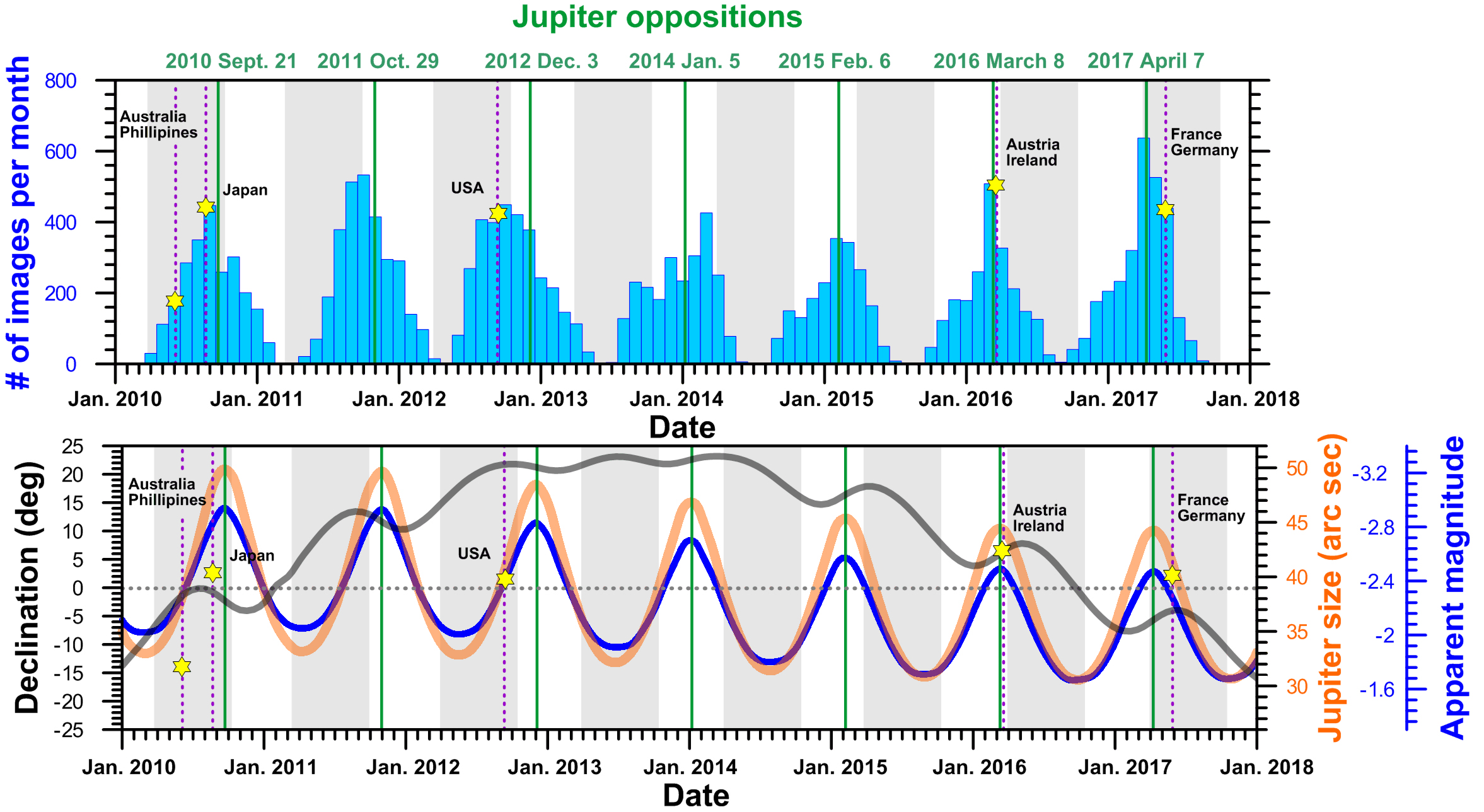}
\caption{Number of Jupiter images per month archived in the PVOL database (top) and observation conditions (bottom) as a function of time. 
Jupiter oppositions are shown with solid green lines and are labeled on the uppermost horizontal axis. 
Dates of Jupiter impacts are shown with vertical dotted magenta lines and yellow stars. The region where the impact was
observed labels each line. Spring and summer months in the north hemisphere are shown in gray, 
and winter and fall in the north hemisphere are shown in white.
Bottom panel: Jupiter declination (left axis and black line) compared with the planet apparent 
size in arcsec (right axis and orange line) compared with its visual magnitude (blue line and axis).
The horizontal dotted gray line shows the zero-declination line.}
\label{fig:PVOL2}
\end{figure*}

\subsection{Continuous flux or meteor showers?}

Several works have predicted the existence of meteor showers on other planets based on the characteristics of 
known comet/meteoroid orbits that cross the nodal points of planets. Most of them considered only the inner planets
(e.g., \citealp{Christou10}), but \cite{Selsis04} presented a study for all solar system planets, including Jupiter and Saturn. 
They found 48 "comet candidates" that could produce meteor showers on Jupiter. In contrast to what happens 
on the inner planets, meteor showers in Jupiter largely overlap in time because of the very long duration of 
close comet passages. We therefore assume that impacts of the class 
detected by amateur astronomers should be distributed randomly in time and not be clustered 
when the Jupiter orbit traverses a cometary tail. We also assume that these impacts 
occur close to the escape velocity of Jupiter of 60 km s$^{-1}$ , in agreement with observations of the SL9 impact \citep{harrington2004book}.

\subsection{Temporal survey of Jupiter from amateur observers}
Understanding the significance of the five flashes requires knowledge about when
the amateur community performs their observations of Jupiter.  Figure ~\ref{fig:PVOL2} shows 
a statistical analysis of Jupiter observations archived at the PVOL database \citep{Hueso10b, Hueso18}. 
This is one of the most popular databases of amateur observations of solar system planets and 
can be searched with very many parameters. In the period from January 2010 to December 2017,
the PVOL database contains 17,643 Jupiter observations that are representative of global trends in 
amateur observations of Jupiter. Each year, the observations cluster more abundantly close to the
opposition of Jupiter. In years where this opposition is close to winter in the north hemisphere, fewer amateurs
are able to observe the planet regularly. This is due to the geographical distribution of most amateur observers. 
About 65\% of all Jupiter observations come from observers in the north hemisphere, with about 21\% of 
observations contributed from $\pm 30^{\circ}$ latitudes and 16\% from south hemisphere latitudes. 
The number of Jupiter observations for the 2010-2011 Jupiter apparition was about 2,400. This number 
reduced by about 25\% in the three Jupiter apparitions in 2013-2015 and increased by about 15\% 
in the latest 2016 and 2017 Jupiter apparitions. 

This trend with Jupiter oppositions helps to explain the gap in the detection of Jupiter flashes in 
the period 2013-2015 when Jupiter opposition resulted in most observers having difficulties to 
find good weather and with generally fewer observations (Figure \ref{fig:PVOL2}). 
Jupiter oppositions in 2018-2022 will occur from May to September, offering increased capabilities of detecting impacts.

\subsection{Statistical interpretation of the flashes}
It is difficult to make statistical analysis of events that have been observed only a few times. 
The results in section 5 provide absolute upper limits and weak lower limits to the observable 
impacts on Jupiter. Here we present different arguments to infer the number of detectable 
impacts on Jupiter. 

\begin{enumerate}
\item An absolute minimum flux of impacts on Jupiter of 0.63 impacts per year is found based on the five flashes detected in eight years (2010-2017). 

\item The statistical analysis from DeTeCt can also be understood as a minimum flux of 4.9 impacts per year. 
Only one of the 11 observers that have successfully found an impact on Jupiter collaborates with this project regularly.
If this fraction is representative of impacts on the planet, then the 4.3 impacts per year could scale up to 52 impacts per year.

\item \cite{Hueso13} gave order-of-magnitude estimates considering the geographical distribution of observers and the number of 
observations per year of the amateur community, inferring between
6-30 detectable impacts per year 
based on the three impacts detected from 2010-2012. We here correct these estimates with an update of 
the number of impacts detected in the period 2010-2017.  We assume that the total surveyed time is given by

\begin{equation}
T=N\cdot t_1\cdot e,
\end{equation}

where $N$ is the total number of images, 17,643 as stated above, $t_1$ is the time accumulated to form each image,  and $e$ is the efficiency for each image
to have enough quality to show an impact. $t_1$ can be from 5 min. to 15 min. since each image is the result of a longer observing
session. The efficiency $e$ in which a video observation can have enough quality so as to show an impact 
was estimated to be from 0.3-0.5 in \cite{Hueso13}. Then $T$ is approximately 18-91 days over an accumulated time 
of eight years. Only 2 of the 11 observers who have found impacts in Jupiter are regular contributors to this database, and we estimate
that the global network of amateur astronomers can be represented by an increase in Jupiter observing time by a factor of 11/2.
Therefore, we consider that the global survey of Jupiter observations by the amateur community can be globally represented by 
a total observation time of 99-500 days obtained over the past eight years. This represents
a global observation efficiency of Jupiter of 3.4-17\%. In this way, the five impacts detected in eight years may scale up to an estimate of 4-18 "detectable" impacts per year.
Even if we were to consider the amateur observations as a "perfect survey" with a detection efficiency of $e=1.0$, the detectability
of impacts would still be limited by the geographical distribution of observers clustered in North and South America, Europe, and Japan-Australia.
This would result in an observing time efficiency of 33\% of the total available time and a minimum number of 2.5 detectable impacts per year.

\item Of the 11 observers that have detected impacts in Jupiter,
2 can be considered as very regular observers 
performing an outstanding number of Jupiter observations every year and participating in several research projects 
(A.W. and C. G.). One of them (A.W.) discovered the debris of an impact in 2009 \citep{SanchezLavega10} and the 
first flash of light the next year. A.W. accumulates 180 hours of Jupiter observations per year, personally looking at 
every video, and half of these data have a quality good enough to visualize a small impact (90 hours per year). This means 
that he has found one (two) impact(s) in an accumulated observing time of 720 (810) 
hours over the past eight (nine) years, where the number in parentheses indicate whether we also consider his initial finding of the 2009 large impact. 
For A.W. alone, this is about 35-40\% of all the time covered by the DeTeCt program. This suggests detectable impacts with a rate 
close to 10-20 impacts per year.

\item The second impact on Jupiter was found only two months after the first. The clustering of these two events 
close in time can be examined with Monte Carlo simulations. 
We simulated impacts considering different impact frequencies in a Monte Carlo simulation representative of eight years, where
for each day, we calculated the random probability of having found an impact.
Detection was examined considering (i) that each day, the planet could only be 
observed 33\% of the time because of the longitudinal distribution of Earth observers that peaks over Europe, North America, and Japan meridians; 
(ii) that observations only cover nine months of a year; 
(iii) that detecting the impact was not possible because of poor weather 50\% of the time; and 
(iv) that detections were not possible because of poor atmospheric seeing 50\% of the time. 
This renders the detection probability of any given impact as 6.3\%.
Two impacts occur with a time difference shorter than three months
in about 50\% of the Monte Carlo simulations, with an impact flux of 5-15 impacts per year.

\end{enumerate}

All in all, we consider that a reasonable estimate of the number of potentially detectable impacts per year in 
Jupiter from objects of 5-20 m size or larger can be on the order of 4-25. 
However, since we only observe the planet nine months every year and we can only observe 
about one half of its surface at any given time, the "detectable" number of impacts 
corresponds to a higher number of objects colliding with Jupiter. This correction means 
that the accumulated flux of objects larger than 5-20 m in diameter that hit 
Jupiter every year is estimated to be 10-65, compared 
with 12-60 from \cite{Hueso13}.

\section{Impact flux on Jupiter}

\begin{figure*}
\centering
\includegraphics[width=15.5cm]{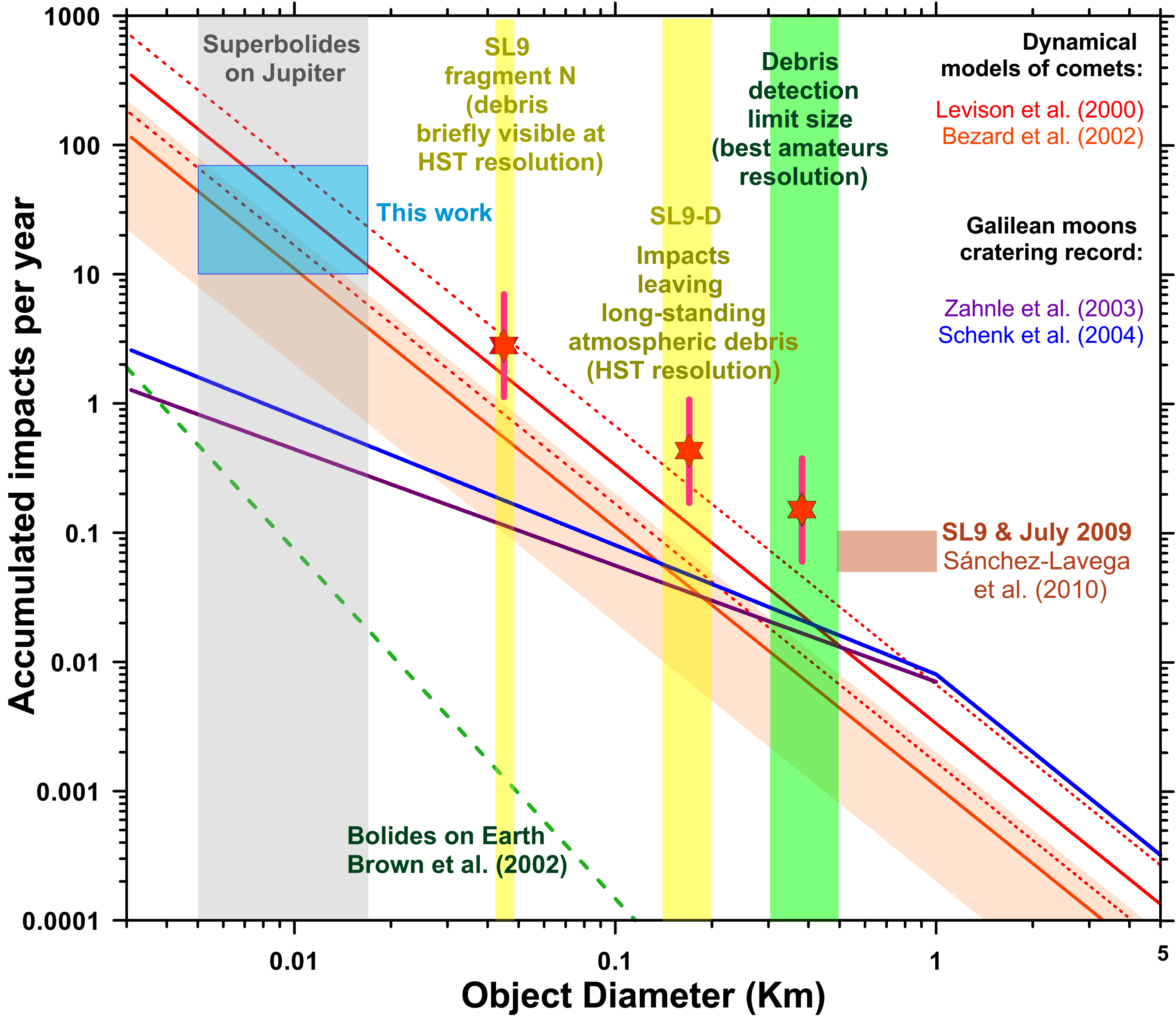}
\caption{Impact rates on Jupiter and Earth compared.  
The vertical gray region represents the sizes of the five bolides, and 
the blue box represents the impact rate of 5-20 m size objects on Jupiter. 
The lines represent impact rates from: (a) dynamical models of comets (red line with 
estimated uncertainties as the dotted line, \citealp {Levison00}; (b) orange line and estimated uncertainties  from corrections to that model introduced by \citealp{Bezard02},
or; (c) the cratering record of Galilean moons (magenta line from \citealp{Zahnle03}; blue line from \citealp{Schenk04}). This is compared with estimates of the impact rate on Earth from \citealp{Brown02}).
The yellow regions indicate the limiting size of objects that might be discovered based on the dark 
debris that they are expected to leave in the atmosphere of Jupiter: The smallest SL9 fragment that 
produced a detectable debris field on HST images (fragment N) and
long-standing debris features associated with objects of 170 m such as SL9 fragment D are highlighted. 
The green vertical region represents objects leaving a debris field that might be detected with amateur 
equipment. The light-brown box represents estimates of impact rates on Jupiter 
from \cite{SanchezLavega10} for 0.5-1.5 km size objects.
Estimates of the impact rate of intermediate-size objects that hit the planet and form
observable debris fields are plotted as red stars. The error bars represent 
an uncertainty factor of 2.5 upward and downward.}Figure updated from \cite{Hueso13}.
\label{fig:ImpactFlux}
\end{figure*}

\subsection{Consequences for the chemichal composition and dust abundance in the Jovian stratosphere}
Based on the flux rate derived in the previous section, the contribution of impacts of this size range 
to the delivery of chemical species and dust to the upper atmosphere
of Jupiter is expected to be on the order 
of $8\times 10^5 - 7\times 10^7$ kg yr$^{-1}$. Recent research on Earth large-size bolides 
shows that the largest fraction of the impacting object is deposited in the atmosphere 
in the form of micrometer dust \citep{Klekociuk15}.
This contribution to exogenous species and dust can be compared with the 
continuous contribution from interplanetary dust particles (IDPs) colliding with Jupiter.
Current models of interplanetary dust fluxes on the giant planets \citep{Poppe16} predict about
$10^{-13}$ g m$^{-2}$ s$^{-1}$  or $1.9\times10^9$ kg yr$^{-1}$. The
dust observations made by the Galileo Dust Detection System (DDS) of 
impact-induced ejecta clouds around the 
Galilean Moons  \citep{Kruger98, Kruger00, Kruger03}
result in estimates of the total mass flux of IDPs on Jupiter of 
$3\times10^{-13}$ g m$^{-2}$ s$^{-1}$ \citep{Sremcevic05}, which
is equivalent to $5.6\times10^9$ kg/yr. 
However, these estimates are probably uncertain by an order of magnitude \citep{Poppe16}.
Thus, the contribution of impacts of the size range such as those discovered in observations 
of Jupiter fireballs contribute about 0.01-3.7 \% of the IDP exogenous material and dust to the upper stratosphere of Jupiter
and are not expected to have a strong impact even on local scales.

\cite{Lellouch02} have analyzed ISO observations and found an upper limit ($8\times10^4$ cm$^{-2}$ s$^{-1}$)  
to the permanent water influx into the stratosphere of Jupiter. \cite{Bezard02} have placed oxygen
influx limits at $(1.5-10)\times10^6$ cm$^{-2}$ s$^{-1}$ 
based on observations of CO.
Various possibilities for the high exogenous CO/H2O ratio at Jupiter have been considered by \cite{Bezard02} 
and are further discussed by \cite{Poppe16} and \cite{Moses17}. The main conclusion is that
favorable production of CO over H2O during IDP ablation is needed to explain this result. 
In the case of the contribution of 5-20 m size impacts, the same conclusions apply, that is to say that the 
incoming water in impacts has to be transformed into CO. The water
influx of $8\times10^4$ cm$^{-2}$ s$^{-1}$ from \cite{Lellouch02} 
is equivalent to a flux of 143,000 kg of water per year. This amount of water could be supplied 
by a single impact of 10 m with 30\% of water if water could be preserved during the impact. 
We conclude that the water molecules ablating from the incoming impacting object must be thermochemically 
converted into CO soon after ablation.

\subsection{Impact flux as a function of impactor size}

\begin{table}
\caption{Summary of impacts on Jupiter}
\centering 
\begin{tabular}{l r r}
\hline\hline
Date & Mass & Reference\\
\hline
1981-03-05     & 11 kg & \citet{Cook81}\\
1994-07-16 to & $1.0 \times 10^9$ Tn   &\citet{Hammel95}\\
1994-07-24    & ~                                        & \citet{harrington2004book}\\
2009-07-19    &  $6.0 \times 10^7$ Tn  & \citet{SanchezLavega10}\\ 
2010-06-03    &  $105-780$ Tn            & \citet{Hueso10b, Hueso13}\\ 
2010-08-20    &  $205-610$ Tn            & \citet{Hueso13}\\ 
2012-09-10    &  $500-950$ Tn            & \citet{Hueso13}\\ 
2016-03-17    &  $403-805$ Tn            & This work\\     
2017-05-26    &  $75-130$ Tn              & This work\\ 
\hline
\end{tabular}
\label{tab:Impact_sizes}
\end{table}

Table \ref{tab:Impact_sizes} summarizes the masses of all impacts observed on Jupiter from the SL9 series of impacts 
down to a very small meteor entering the atmosphere of Jupiter
that was observed by Voyager 1 \citep{Cook81}. 
Other plausible but unconfirmed collisions with Jupiter have been proposed, such as a dark spot 
on Jupiter observed by Cassini in 1690 with morphological characteristics and behavior similar to SL9 debris features 
\citep{Tabe97} that were proposed to be caused by an impact with a 600 m object \citep{Zahnle03}.

Figure \ref{fig:ImpactFlux} presents an update of results presented in \cite{Hueso13} of our 
understanding of the current flux of impacts on Jupiter. The
results from this work are very similar to our
previous analysis. The deduced impact rate for 5-20 m size objects compared with estimates of greater impacts 
from \citet{SanchezLavega10} can be interpolated to predict plausible impact rates of 
intermediate-size objects, as shown with stars in Figure \ref{fig:ImpactFlux}. Error bars
here represent uncertainties of 2.5 higher or lower from the overall uncertainty of the impact
rate of small-size impacts (blue box in Figure \ref{fig:ImpactFlux}). This is compared with 
expectations from dynamical models of comets prone to Jupiter collisions 
\citep{Levison00} and the cratering record on the Galilean moons \citep{Zahnle03, Schenk04}. 
Predictions of impact rates from this study are comparable to the upper estimates of impact rates from 
\citet{Levison00}.
Our results for small-size impacts clearly depart from impact rates estimated from cratering of the Jovian moons, 
which for small impacts are dominated by young craters on Europa \citep{Schenk04}.

\subsection{Searches of debris left by impacts}

We now focus on predictions based on Figure \ref{fig:ImpactFlux} of the plausible impact rate
of larger objects that might be detected as an intense flash and might
leave an observable trace in the atmosphere. The detectability of these 
impacts depends not only on the impact frequency in Figure \ref{fig:ImpactFlux},  
it also depends on their size and the biases associated with their detection. 
Predicted impact rates in the planet and their detectatibility 
are summarized in Table \ref{tab:Impact_predictions} and are
discussed below.

\begin{table}
\caption{Predictions of impacts per year on Jupiter that leave observable debris fields, and their detectability}\centering 
\begin{tabular}{l c c c c c}
\hline\hline
Size (m)     &    Mean           &     Min            & Max                & Detectability \\
                &   (yr$^{-1}$)   &   (yr$^{-1}$)   &  (yr$^{-1}$)    &   (yr)\\
\hline 
45             &      2.8    &      1.1      &      7.0    &  0.4-2.6  \\
170           &      0.43   &      0.17    &      1.1    &  2-12      \\
380           &      0.15   &      0.06    &      0.37  &  6-30    \\
\hline
\end{tabular}
\begin{flushleft}
Note: The impact frequency is the number of impacts that we estimate to occur on Jupiter every year. 
Their detectability (the mean number of years between observable impacts of a given size and larger) is 
affected by observational biases, as discussed in detail in section 7.3.
\end{flushleft}
\label{tab:Impact_predictions}
\end{table}

\begin{itemize}

\item Large superbolides. Voyager and Cassini observed Jupiter at high resolution over a period of at least three months. For Cassini, a global coverage at spatial 
resolutions of about 140 km/pix or better was acquired for at least 15 days \citep{Porco03, Salyk06}. Objects of 5-20 m hitting the planet
with the flux rate deduced from this work would give a non-negligible probability (0.5-3 impacts over 15 days) to have occurred in the 
course of this Cassini 15-day window. Higher resolutions over particular regions were acquired for another 30 days. 
Thus, small debris fields in the methane band and ultraviolet images where the debris maximizes its contrast 
might exist in the Cassini imaging data. On Earth, many satellites have observed high-atmosphere debris associated with impacts. The best case are
the satellite observations of the Chelyabinsk impact, which was observed at a variety of spatial resolutions 1-10 km/pix for at least 3 hr \citep{Miller13}. 
This impact was similar in energy to the Jupiter impacts we discussed
here. Although other missions have imaged Jupiter (Pioneers, New Horizons, and Juno), the number of images 
from these missions is too low to merit a specific analysis, but for Cassini and the Voyagers, a search for tiny- and small-debris fields
might place an important constraint in the rate of impacts on the planet.

\item 
Small impacts. Based on this analysis, objects of 45 m or larger that leave a short-lived debris field that is only observable with large telescopes
may impact Jupiter once every 0.36 years with uncertainties from 0.14 to 0.9 years. 
Since they can impact on the far side of the planet or in months when Jupiter is not observable, 
even a perfect survey of  impacts on Jupiter could only find these events once per year with estimated uncertainties 
from 0.4 to 2.6 years. HST observations of the fragment N impact site did not allow determining for how long
debris from this impact might be observable. The visibility of such an impact may also depend on its latitudinal location and on dissipation effects
such as local wind shear. A careful examination of archived HST images of Jupiter acquired since 1991 (the date of the first Jupiter 
observation) and a search for tiny dark spots in the visible or that are bright in methane may place additional constraints on the impact rate 
on Jupiter or serve to lower the impact rate deduced from small-size flashing impacts. This is a non-trivial effort because archived
HST Jupiter images encompass more than 350 target names, several instruments, tens of filters, 
and observing programs that range from global coverage to snapshots, and the debris field might be within the limit of detectability in most filters. 
Additionally, small impact debris fields would not be observable in subpolar latitudes.

\item 
Intermediate-size impacts. Objects larger than 170 m that leave a debris field that might be observable with 
small telescopes over weeks and months with professional telescopes may occur once every 
2.3 years with estimated uncertainties from 1 to 6 years. Their detectability is difficult 
to ascertain because these events might be observed in only about half of each year when Jupiter is 
well placed in the night sky for astronomical observation. Therefore a timescale of 2-12 years seems reasonable for the detectability of these events.

\item 
Large impacts. Objects larger than 380 m that are able to leave a debris field that is observable in standard amateur images over weeks and 
during months in observations with professional telescopes may only occur once every 7 years with uncertainties 
from 3 to 16 years. Again, since Jupiter is only observable nine months every year and the quality of observations is a function of the proximity 
to Jupiter opposition, these events may be discovered by amateurs about once every 6-30 years, 
similarly to the timescale separation between the SL-9 collision and the 2009 impact.
\end{itemize}

\section{Conclusions}

The most recent two impacts on Jupiter in March 2016 and May 2017 had masses of 
310-620 Tn and  75-130 Tn, respectively, with sizes ranging from 4.1 m to 17 m 
for object densities from 2.0 to 0.25 g cm$^{-3}$. These masses are comparable to 
previous impacts on Jupiter that have also been found in flashes in video observations of the planet.

The cumulative impact rate of objects of this size range or larger is predicted to lie in the range of 
10-65 per year, with only 4-25 impacts per year being observable in a perfect survey of flashes
because of the way they are distributed over Jupiter's visible and far sides and the Jupiter 
observation period per year.

The overall impact rate for Jupiter according to this work is 
similar to the high impact rate limit of \cite{Levison00} and compatible with the modifications
discussed by \cite{Bezard02}. Although significant uncertainties exist, the observations of impact flashes on Jupiter disprove
the low impact rate predicted from the cratering record on Europa that was discussed by \cite{Schenk04}.

5-20 m size objects impacting Jupiter are expected to be detected yearly
in the next Jupiter oppositions because of the improved observing conditions and the 
availability of software tools.

Future observations of impacts with the modern cameras
currently available to the amateur community could be obtained at 60 fps or higher
with a good signal-to-noise ratio. For an energetic impact like those of 2012 and 2016, this
may allow exploring the fragmentation history of impacting objects, opening the possibility 
of studying the nature of the impacting object (stony, metallic, icy compact, or icy porous).

The accumulated effect of these impacts on the chemistry of the upper stratosphere of the 
planet is negligible when compared with other sources of exogenous chemicals, such as
interplanetary dust particles and giant impacts.

A “large flash” comparable to fragment N of SL9 leaving an observable 
debris field at the limit of spatial resolution with HST, VLT, or other large telescopes might occur on Jupiter with a typical timescale of once every 2-11 months. A perfect
observational survey of bright flashes would find these powerful flashes about once per year. 
A dedicated careful examination of all HST observations of Jupiter obtained since 1991 that would search for small dark spots on visible images and
bright spots on methane images mighthelp to reduce the factor of 6 uncertainty on the impact rate from this work. A similar search
for smaller impact debris from Voyager and Cassini images seems worthwhile, based on this study.

An extremely intense  flash leaving a standing debris field in the atmosphere of Jupiter
that could be observed with HST or large ground-based telescopes over weeks 
might occur on Jupiter once every 1-6 years. Regular ground-based observations
of debris fields on Jupiter might detect these events about once every 2-12 years and more 
efficiently than a survey of flashes.

Impacts leaving a debris field that would be observable with amateur equipment might 
occur on Jupiter once every 3-16 years and might be observable once every 6-30 years 
when accounting for the time of the year when Jupiter can be observed at 
high resolution by amateur astronomers.

Future observations will find increasingly smaller impacts as the technology improves. 
Specific searches in Voyager, Cassini, and HST images may contain small-debris fields that were
not detected at the time of the acquisition of these observations. If these "small impact scars" are detected,
they will largely constrain the impact rate on Jupiter. The JUICE mission to Jupiter 
may also discover flashing impacts on the planet that are caused by objects of much smaller
sizes through the long surveys of the Jupiter night side that
are currently planned for studies 
of the Jovian magnetosphere and the deep lightening activity \citep{Grasset13}.

\begin{acknowledgements}
We are very grateful to Thomas Riessler for permission to work on his recording of the May 2017 impact. We are also 
grateful to Emmanuel Lellouch for a detailed and constructive review of this research. 
We thank C. Go, M. Tachikawa, K. Aoki, M. Ishimaru, D. Petersen, and G. Hall from their observation of impacts in Jupiter 
as well as I. Tabe for help in communicating with Japanese amateur astronomers. We also thank
Sebastian Voltmer for providing early information from Gerrit Kernbauer's video detection and image processing of this video,
and Dan Fischer for his help in finding German-speaking observers that had observed the May 2017 impact.
We are also very grateful to the ensemble of amateur astronomers running DeTeCt on their video observations of Jupiter, and
to the large community of observers providing Jupiter observation data through PVOL and ALPO Japan.
R.H. and A.S.L. were supported by the Spanish project AYA2015-65041 (MINECO/FEDER, UE), 
Grupos Gobierno Vasco IT-765-13 and UPV/EHU UFI11/55. This work has been supported by the
Europlanet 2020 Research Infrastructure.  Europlanet 2020 RI has received funding from the 
European Union's Horizon 2020 research and innovation programme under grant agreement No 654208.

\end{acknowledgements}

%
%

\bibliographystyle{aa} 
\bibliography{bibliography_AA} 

\end{document}